\documentclass{article}
\usepackage{spconf,amsmath,amssymb,graphicx,hyperref}

\title{GRAM BARCODES FOR HISTOPATHOLOGY TISSUE TEXTURE RETRIEVAL}

\name{Shalev Lifshitz$^{\star \dagger}$, Abtin Riasatian$^{\star}$, H.R. Tizhoosh$^{\star\ddagger}$}
  
\address{$^{\star}$Kimia Lab, University of Waterloo, Canada \\ $^{\dagger}$Computer Science, University of Toronto, Canada \\
$^{\ddagger}$Vector Institute, Toronto, Canada}

\begin{document}

\maketitle

\begin{abstract}
Recent advances in digital pathology have led to the need for Histopathology Image Retrieval (HIR) systems that search through databases of biopsy images to find similar cases to a given query image. These HIR systems allow pathologists to effortlessly and efficiently access thousands of previously diagnosed cases in order to exploit the knowledge in the corresponding pathology reports. Since HIR systems may have to deal with millions of gigapixel images, the extraction of compact and expressive image features must be available to allow for efficient and accurate retrieval. In this paper, we propose the application of \textbf{Gram barcodes} as image features for HIR systems. Unlike most feature generation schemes, Gram barcodes are based on high-order statistics that describe tissue texture by summarizing the correlations between different feature maps in layers of convolutional neural networks. We run HIR experiments on three public datasets using a pre-trained VGG19 network for Gram barcode generation and showcase highly competitive results.
\end{abstract}


\section{Introduction}
\label{sec:intro}

The recent rise in popularity of digital pathology and whole slide imaging has generated a plethora of histopathology image data. As a result, interest in Histopathology Image Retrieval (HIR) has also increased since it allows pathologists to conveniently find similar images to a query image in order to exploit the available knowledge of previously diagnosed cases. Many HIR systems extract features of images and calculate a similarity score for two images by applying a similarity metric on their corresponding features. Thus, HIR systems can retrieve the most similar images to a query image by generating features for all images in a database and retrieving the cases that are determined to be the most similar. There are, however, some challenges preventing HIR systems from achieving practical implementation: 1) Large amounts of representative data are needed to be able to retrieve similar images, 2) Retrieval times are generally slow and impractical due to the large size of images and their corresponding features, 3) Retrieval accuracy is often not sufficiently high because feature extraction algorithms may not account for tissue polymorphism. Therefore, innovative methods for extracting compact and expressive features of medical images must be proposed to advance the field.

Using Convolutional Neural Networks (CNNs) to extract image features has been very useful for HIR systems, but research using CNNs for feature extraction has mainly focused on extracting first-order features, i.e.,  ``pooling or encoding methods are adopted on feature maps directly to produce compact image [features]'' \cite{zhao2017gram}. Exploration of high-order methods for generating features in HIR systems, such as determining ``the dependencies between different channels in the same [CNN] layer'' \cite{zhao2017gram},  might yield extraction methods that generate more compact and expressive features. The literature on high-order feature extraction for image retrieval is quite sparse. There have been few works with some relation to high-order feature extraction for HIR, specifically.

In this paper, we investigate the application of \textit{Gram barcodes} as image features for HIR systems. Gram barcodes are spatially invariant high-order representations of images that are built off of the Gram matrix proposed by Gatys et al. in \cite{gatys2015texture,gatys2015neural}, where it was used as a measure of image texture and style. Gram barcodes, as we propose, offer the potential to accurately measure and efficiently compare tissue textures between histopathology images by describing the relationship between different feature maps in the same CNN convolution layer. We conducted HIR experiments with three public datasets using a pre-trained VGG19 network \cite{simonyan2014very} for Gram barcode generation and showcase highly competitive results. The associated code for this paper can be found at \url{https://github.com/Shalev-Lifshitz/GramBarcodes}.

\section{Background}
\label{sec:background}

\subsection{Histopathology Image Analysis with Deep Learning}
Deep learning has shown great promise for feature extraction in HIR systems. However, HIR systems have mainly used first-order features produced by CNNs to represent images and determine their similarity \cite{hegde2019similar,kalra2020pan,kumar2018deep}. Hegde et al. \cite{hegde2019similar} introduced SMILY, a content-based image retrieval tool for histopathology images. Kalra et al. \cite{kalra2020pan} performed HIR on 20 million image patches (1000 $\times$ 1000 pixels) and report superior results for various tumour types. Kumar et al. \cite{kumar2018deep} proposed deep barcodes as features for HIR, since binary features allow for faster image retrieval.

Deep learning has also been used for various other tasks in histopathology image analysis \cite{kensert2019transfer,xu2017Large,rakhlin2018deep}. Kensert et al. \cite{kensert2019transfer} used pre-trained CNNs to ``predict cell mechanisms of action in response to chemical perturbations for two cell profiling datasets from the Broad Bioimage Benchmark Collection" \cite{kensert2019transfer}. They obtained high predictive accuracy between $94\%$ and $97\%$. Xu et al. \cite{xu2017Large} used AlexNet to perform classification, segmentation, and visualization for large-scale histopathology tissue images (brain tumor and colon cancer datasets). Pre-trained ResNet50, InceptionV3, and VGG16 networks were used by Rakhlin et al. \cite{rakhlin2018deep} to classify breast histology microscopy images. They reported $87.2\%$ accuracy for a 4-class classification task and $93.8\%$ accuracy for a 2-class classification task to detect carcinomas.

\subsection{Gram Matrix}
In 2015, Gatys et al. \cite{gatys2015texture,gatys2015neural} defined the Gram matrix as a spatially invariant summary statistic that can effectively describe the texture, or style, of an image by analyzing the correlations between different feature maps within the same CNN convolution layer. Note that texture and style are used interchangeably throughout this work.

\begin{figure}[t]
\centering
\includegraphics[scale=0.25]{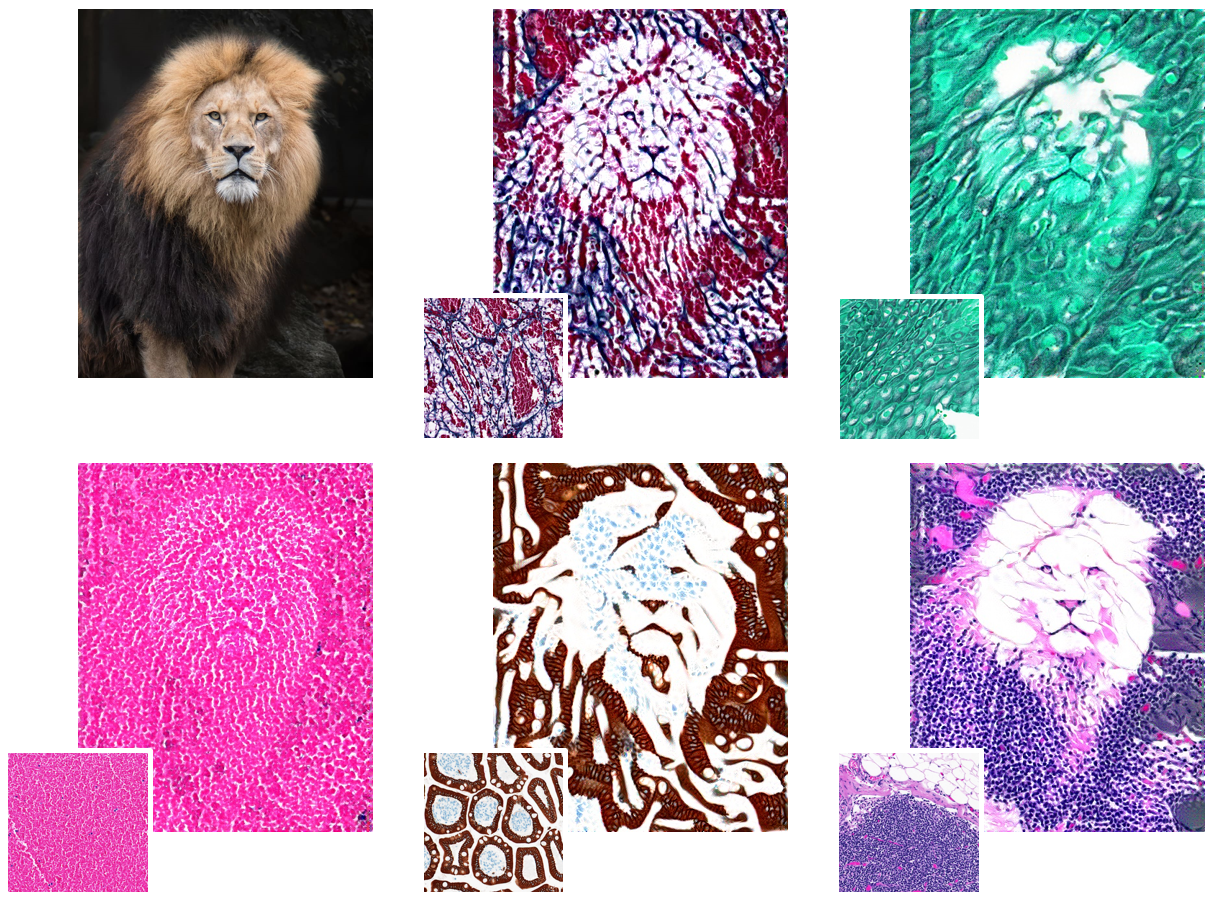}
\caption{To showcase the Gram matrix's ability to describe histopathological tissue texture, we transferred the texture of histopathology images from the KimiaPath24 dataset \cite{Babaie_2017_CVPR_Workshops} onto a content image (\textit{lion}) using the method outlined in \cite{gatys2015neural}. 
\label{fig:cell_style}}
\end{figure}

In \cite{gatys2015neural}, Gatys et al. combined the content of one image (content image) with the style of another image (style image) by simultaneously minimizing the content loss and style loss between a randomly generated white noise input image and the content and style images, respectively. The style loss is defined as the mean squared error between the Gram matrices of the input image and style image. Figure \ref{fig:cell_style} showcases the ability of the Gram matrix to describe histopathological tissue texture by using the algorithm outlined in \cite{gatys2015neural} to transfer the texture of various histopathology images from the KimiaPath24 dataset \cite{Babaie_2017_CVPR_Workshops} onto a content image. For the purpose of creating the figure, the input image to the algorithm was initialized as a copy of the content image.

Following these two works by Gatys et al., the Gram matrix was used in a number of works as a representation of image texture and style \cite{holden2017fast,johnson2016perceptual}. The Gram matrix was also used to perform style transfer on videos \cite{ruder2016artistic}. Image style classification systems have made use of the Gram matrix to describe image style \cite{chu2016deep} and Yu et al. \cite{yu2019speaking} used the Gram matrix to recognize the apparent personality of a subject based on their speaking style. Gram matrices have recently been used as features for image retrieval  \cite{zhao2017gram,matsuo2016cnn,wang2018deep}. However, to our knowledge, there has been no work directly related to binary Gram matrices as features for HIR, specifically.

\section{Methods}
\label{sec:methods}

\textbf{Gram Barcode Generation --} The Gram matrix $\textbf{G}^l$ is a summary statistic that is generated by passing an image through a CNN and computing the correlations between the feature maps at a convolution layer $l$. To compute the Gram matrix $\textbf{G}^l$, the feature maps of layer $l$ are first stored in a matrix $\mathbf{F}^l \in \mathbb{R}^{N_l \times M_l}$, where $N_l$ is the number of feature maps in layer $l$ and $M_l$ is the size of a single feature map when vectorized. Each element $\mathbf{G}^l_{ij}$ of Gram matrix $\mathbf{G}^l \in \mathbb{R}^{N_l \times N_l}$ is defined as the inner product between vectorized feature maps $i$ and $j$ in layer $l$ \cite{gatys2015neural}:
\begin{equation} 
\mathbf{G}_{ij}^l = \sum\limits_{k} \mathbf{F}_{i k}^l \mathbf{F}_{j k}^l.
\end{equation}
Importantly, when the inner product is calculated, the resulting element $\mathbf{G}_{ij}^l$ loses all spatial information. Thus, the Gram matrix is spatially invariant and does not describe local structures, but rather describes stationary textures and patterns as the correlations between the feature maps in layer $l$ \cite{gatys2015texture}.

A set of Gram matrices \{$\mathbf{G}^1, \mathbf{G}^2, ..., \mathbf{G}^L$\} can be calculated for any $L$ specified convolution layers in a CNN. To convert a set of Gram matrices into a Gram barcode, we performed three additional steps. First, since a Gram matrix is symmetric, we remove the lower or upper triangle of each Gram matrix to reduce redundant computation during retrieval. Then, each Gram matrix in the set is vectorized and concatenated, creating a larger Gram vector $\mathbf{v}$. Finally, to allow for efficient retrieval, $\mathbf{v}$ is binarized using the global median threshold $\widetilde{v}$ (the median of $\mathbf{v}$) to create the final Gram barcode $\mathbf{b}$. Specifically, each element $\mathbf{b}_i$ is given by
\begin{equation}
\displaystyle \mathbf{b}_i = \begin{cases} 
1 & \text{$\mathbf{v}_{i}\geq \widetilde{v}$} \\
0 & \text{$\mathbf{v}_{i} < \widetilde{v}$} 
\end{cases}
\end{equation}

\textbf{Image Retrieval with Gram Barcodes --} In HIR systems, a \textit{query image} is specified as the input image for a retrieval algorithm, which aims to retrieve the images in a database that are most similar to the query image. In general, to perform a retrieval, features generated for the query image are matched against the features of all images in the database, and the images with the most similar features (to the query image features) are retrieved. Importantly, this database is pre-indexed, which means that the features of all images are generated and saved before any queries are made. In this way, features for all images in the database do not need to be computed each time the system is queried. 

In this paper, we perform HIR with Gram barcodes as features. The similarity between the Gram barcode of the query image and those of all other images in the database is determined by computing their logical XOR difference (i.e., Hamming distance), which has been shown to be an effective metric for determining image similarity in HIR systems \cite{kumar2018deep}. The Gram barcodes with the smallest XOR distance to that of the query image are determined to represent the most similar images and the top-$n$ most similar images are retrieved. 

\section{Datasets}
\label{sec:datasets}

We perform experiments on three publicly available datasets that contain histopathology images: the KimiaPath24 dataset, a Colorectal Cancer (CRC) dataset, and an Endometrial Cancer (EMC) dataset. 

\textbf{KimiaPath24 dataset --} Introduced  by  Babaie et  al. \cite{Babaie_2017_CVPR_Workshops}, KimiaPath24 is a small dataset containing 24 Whole Slide Images (WSIs) of different tissue textures with 1,325 test patches of size 1000 $\times$ 1000 pixels (0.5$\mu m$ $\times$ 0.5$\mu m$) that were carefully extracted from all 24 WSIs ``with special attention to textural differences" \cite{Babaie_2017_CVPR_Workshops}. Approximately 27,000 to over 50,000 training patches can be extracted from the 24 WSIs, depending on the selection of homogeneity and overlap for every WSI \cite{Babaie_2017_CVPR_Workshops}.

\textbf{CRC dataset --} Kather et al. \cite{kather2016multi} created this dataset to assess the performance of various texture descriptors and classifiers. It consists of 5,000 images (150 $\times$ 150 pixels) of human CRC divided across 8 different classes (each containing 625 images): tumour epithelium, simple stroma, complex stroma, immune cells, debris, normal mucosal glands, adipose tissue, and background patches. 

\textbf{EMC dataset --} Sun et al. \cite{sun2019computer} introduced this dataset to compare the classification ability of a deep learning method (HIENet) to three experienced pathologists, as well as five other CNN-based classifiers. The dataset includes  almost 3,300 image patches (640 $\times$ 480 pixels) from almost 500 endometrial specimens in four different classes of endometrial tissue (each with a varying number of images): normal, endometrial polyp, endometrial hyperplasia, and endometrial adenocarcinoma. The patches are extracted from 20$\times$ or 10$\times$ magnified WSIs and, although all slides have been prepared and scanned at the same hospital, considerable stain variation can be observed across endometrium patches.

\subsection{Evaluation Metrics}
To create comparable results, we evaluate each dataset using the metrics that were put forth by its creators.

\textbf{KimiaPath24 --} There are a total of $n_{tot} = 1325$ patches $P^i_s$ that belong to 24 sets $\Gamma_s = \{P^i_s \mid s \in S, i = 1, 2, ..., n_{\Gamma_s}\}$ with $s = 0, 1, 2, ..., 23$ ($n_{\Gamma_s}$ being the number of test patches for each WSI). Where $R$ is the set of retrieved images for any experiment, the \textbf{patch-to-scan accuracy} $\eta_p$ is defined as
\begin{equation}
    \eta_p = \frac{\sum_{s \in S} | R \cap \Gamma_s |}{n_{tot}}
\end{equation}
and the \textbf{whole-scan accuracy} $\eta_W$ is defined as 
\begin{equation}
    \eta_W = \frac{1}{24} \sum_{s \in S} \frac{| R \cap \Gamma_s |}{n_{\Gamma_s}} .
\end{equation}
In addition, the total accuracy is defined as $\eta_{total} = \eta_p \times \eta_W$.

\textbf{CRC and EMC datasets --} Both datasets are evaluated using the well-known evaluation metrics accuracy, sensitivity, specificity, and area under the curve (AUC). Specifically:
\begin{align}
    \textrm{Accuracy} &= \frac{TP + TN}{TP + TN + FP + FN} \\
    \textrm{Sensitivity} &= \frac{TP}{TP + FN} \\
    \textrm{Specificity} &= \frac{TN}{TN + FP}
\end{align}
where TP, TN, FP, and FN are the number of true positives, true negatives, false positives, and false negatives, respectively, for any experiment. To compute these metrics, we predict the class of each test image and determine whether the prediction is a TP, TN, TP, or FN. Also, since the datasets we address contain more than two classes, when considering the class of a query image, the class of the query image is defined as positive and all other classes are defined as negative.

\begin{table*}[]
    \centering
    \resizebox{\textwidth}{!}{
    \begin{tabular}{|l|c|c|c|c|c|c|c|c|c|c|c|}
    \hline
    Layers & \multicolumn{3}{|c|}{KimiaPath24} & \multicolumn{4}{|c|}{Colorectal Cancer (CRC) Dataset} & \multicolumn{4}{|c|}{Endometrial Cancer (EMC) Dataset} \\
    \hline
    & $\eta_p$ & $\eta_W$ & $\eta_{tot}$ & Accuracy & Sensitivity & Specificity & AUC & Accuracy & Sensitivity & Specificity & AUC \\
    \hline
    1 & 81.74 & 82.21 & 67.20 & 91.28 ± 0.70 & 91.28 ± 0.70 & 98.75 ± 0.10 & 0.9779 ± 0.0027 & 58.14 ± 1.43 & 55.69 ± 1.32 & 85.37 ± 0.49 & 0.7701 ± 0.0096 \\
    \hline
    2 & 82.19 & 81.97 & 67.37 & 90.18 ± 1.00 & 90.18 ± 1.00 & 98.60 ± 0.14 & 0.9737 ± 0.0052 & 57.57 ± 2.80 & 55.31 ± 2.74 & 85.08 ± 0.93 & 0.7654 ± 0.0172 \\
    \hline
    3 & 89.81 & 90.47 & 81.25 & 92.02 ± 1.04 & 92.02 ± 1.04 & 98.86 ± 0.15 & 0.9808 ± 0.0043 & 66.81 ± 1.28 & 65.02 ± 1.27 & 88.21 ± 0.42 & 0.8413 ± 0.0112 \\
    \hline
    4 & 90.49 & 91.01 & 82.36 & 93.38 ± 1.00 & 93.38 ± 1.00 & 99.05 ± 0.14 & 0.9823 ± 0.0038 & 69.02 ± 2.55 & 67.38 ± 2.51 & 88.93 ± 0.88 & 0.8591 ± 0.0187 \\
    \hline
    5 & 92.38 & 93.26 & 86.15 & 94.28 ± 0.59 & 94.28 ± 0.59 & 99.18 ± 0.08 & 0.9867 ± 0.0020 & 73.77 ± 1.68 & 72.35 ± 1.48 & 90.54 ± 0.58 & 0.8870 ± 0.0095 \\
    \hline
    1,5 & 92.15 & 92.92 & 85.62 & 94.58 ± 0.59 & 94.58 ± 0.59 & 99.23 ± 0.08 & 0.9874 ± 0.0020 & 73.80 ± 1.53 & 72.48 ± 1.30 & 90.60 ± 0.53 & 0.8896 ± 0.0077 \\
    \hline
    2,5 & 92.45 & 93.29 & 86.25 & 94.70 ± 0.53 & 94.70 ± 0.53 & 99.24 ± 0.08 & 0.9870 ± 0.0018 & 73.89 ± 1.64 & 72.62 ± 1.46 & 90.62 ± 0.57 & 0.8896 ± 0.0081 \\
    \hline
    3,5 & 93.06 & 93.49 & 87.00 & 94.88 ± 0.75 & 94.88 ± 0.75 & 99.27 ± 0.11 & 0.9881 ± 0.0028 & \textbf{74.92 ± 1.69} & \textbf{73.67 ± 1.43} & \textbf{90.99 ± 0.57} & \textbf{0.8953 ± 0.0072} \\
    \hline
    4,5 & 92.15 & 92.44 & 85.18 & 94.88 ± 0.95 & 94.88 ± 0.95 & 99.27 ± 0.14 & 0.9877 ± 0.0027 & 74.23 ± 1.99 & 72.87 ± 1.67 & 90.79 ± 0.63 & 0.8961 ± 0.0106 \\
    \hline
    2,3,5 & \textbf{93.21} & \textbf{94.19} & \textbf{87.79} & 95.08 ± 0.54 & 95.08 ± 0.54 & 99.30 ± 0.08 & 0.9883 ± 0.0025 & 74.47 ± 2.17 & 73.41 ± 1.88 & 90.87 ± 0.73 & 0.8957 ± 0.0059 \\
    \hline
    1,2,3,5 & 93.21 & 94.08 & 87.69 & 95.08 ± 0.48 & 95.08 ± 0.48 & 99.30 ± 0.07 & 0.9886 ± 0.0021 & 74.56 ± 1.84 & 73.49 ± 1.48 & 90.92 ± 0.61 & 0.8972 ± 0.0046 \\
    \hline
    1,2,3,4,5 & 92.91 & 93.50 & 86.87 & \textbf{95.34 ± 0.76} & \textbf{95.34 ± 0.76} & \textbf{99.33 ± 0.11} & \textbf{0.9881 ± 0.0035} & 73.89 ± 1.85 & 72.66 ± 1.71 & 90.66 ± 0.61 & 0.8945 ± 0.0112 \\
    \hline
    \end{tabular}}
    \caption{Results for all datasets with multiple combinations of the VGG19 layers \lq conv1\_1\rq, \lq conv2\_1\rq, \lq conv3\_1\rq, \lq conv4\_1\rq, and \lq conv5\_1\rq \ being reported (denoted in the ``Layers'' column by the numbers 1, 2, 3, 4, and 5, respectively). Metrics for the KimiaPath24 dataset were computed on the provided training and testing sets. Metrics for the CRC and EMC datasets were obtained using 5-fold cross validation. All values, except AUC, are given in percentages. Notice that results tend to improve as we use deeper layers and employ combinations of layers to generate gram barcodes.}
    \label{tab:results}
\end{table*}

\section{Experiments}
\label{sec:experiments}

\begin{figure}[htb]
\centering
\includegraphics[scale=0.16]{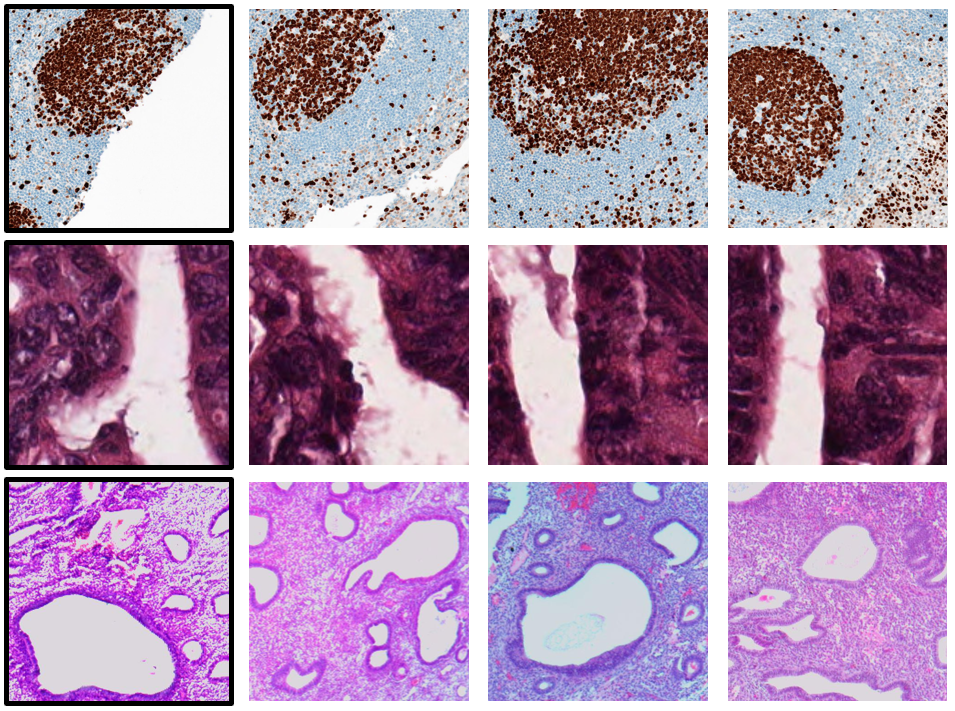}
\includegraphics[scale=0.16]{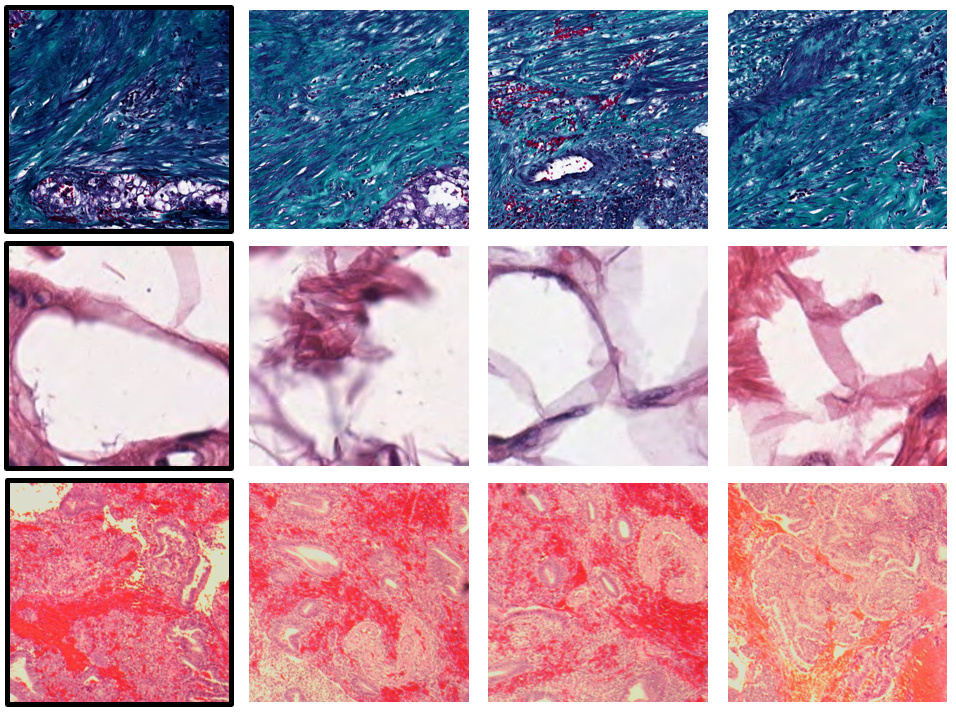}
\caption{HIR examples of successful (left 4 columns) and unsuccessful (right 4 columns) class predictions for the KimiaPath24 dataset (first row), CRC dataset (second row), and EMC dataset (third row). The query images have a black outline.} 
\label{fig:success_cases}
\end{figure}

All experiments were performed using a pre-trained VGG19 network \cite{simonyan2014very} (ImageNet weights) for Gram barcode generation. The results obtained from using various combinations of VGG19 layers for Gram barcode generation are described in Table \ref{tab:results} and examples of successful and unsuccessful retrievals for each dataset are showcased in Fig. \ref{fig:success_cases}. It is noteworthy to mention that the accuracies in Table \ref{tab:results} tend to increase as we use deeper VGG19 layers for Gram barcode generation. Further, combining multiple layers boosts performance. 

In our experiments, the predicted class of any query image is the mode of the classes of the \textbf{top-3} most similar images retrieved by our algorithm. If there is no mode, the class of the most similar image (the top-1 image) is predicted to be the class of the query image.

\textbf{KimiaPath24 dataset --} We achieved a top $\eta_{total}$ score of 87.79\% ($\eta_p =$ 93.21\%,~ $\eta_w =$ 94.19\%) on the provided training and testing sets using VGG19 layers \lq conv2\_1\rq, \lq conv3\_1\rq, and \lq conv5\_1\rq \ for Gram barcode generation. The highest reported $\eta_{total}$ score we could find in the literature was 94.95\% ($\eta_p =$ 97.89\%,~ $\eta_w =$ 97.00\%) by Yang et al. \cite{yang2020deep} and, as far as we could tell, our method is the only other work that has achieved $\eta_p$ and $\eta_w$ scores above 90\% for this dataset.

\textbf{CRC dataset --} We performed 5-fold cross validation on the CRC dataset and achieved a top accuracy of 95.34\% $\pm$ 0.76\% using VGG19 layers \lq conv1\_\rq, \lq conv2\_1\rq, \lq conv3\_1\rq, \lq conv4\_1\rq, and \lq conv5\_1\rq \ for Gram barcode generation. Our method obtains state-of-the-art performance when compared to other retrieval algorithms that have been employed on the same dataset. Specifically, the best retrieval method we could find reported an accuracy of 92.98\% \cite{ebrahimian2020class}. However, our method falls short of the \textit{non-retrieval} based method employed by Nanni et al. \cite{nannia2020ensemble}, who train an ensemble of neural networks for classification and report 97.60\% accuracy.

\textbf{EMC dataset --} We performed 5-fold cross validation on the EMC dataset and achieved a top accuracy of 74.92\% $\pm$ 1.69\% using VGG19 layers \lq conv4\_1\rq \ and \lq conv5\_1\rq \ for Gram barcode generation. Our result is comparable to state-of-the-art retrieval and non-retrieval based methods employed on the same dataset. Specifically, the best method we could find in the literature was a retrieval algorithm reporting 78.59\% accuracy \cite{ebrahimian2020class}.

\section{Conclusions}
\label{sec:conclusions}

We introduced the notion of the \textit{Gram barcode} and investigated its application as image features for HIR. The Gram barcode is a high-order summary statistic that describes histopathology image tissue texture by summarizing the correlations between feature maps in CNN convolution layers. Gram barcodes show highly competitive accuracy levels when applied on 3 public histopathology datasets and, when compared to other retrieval algorithms in the literature, Gram barcodes show state-of-the-art performance for the Colorectal Cancer dataset \cite{kather2016multi}, specifically. Furthermore, Gram barcodes can be calculated using various combinations of CNN convolution layers to boost performance. Gram Barcodes address important barriers to the adoption of HIR systems (slow retrieval times, large features, and low accuracies) by providing more efficient retrieval, compact features, and excellent accuracies on all three datasets. Next steps might include using different CNN architectures for Gram barcode generation and applying dimensionality reduction techniques on the Gram barcodes to decrease retrieval time.

\bibliographystyle{IEEEbib}
\bibliography{myrefs}

\end{document}